\documentclass[aip,jcp,reprint,longbibliography,nofootinbib]{revtex4-2}

\usepackage{amsmath,amssymb,mathtools,bm}
\usepackage{graphicx}
\usepackage{booktabs}
\usepackage{array}
\usepackage{dcolumn}
\usepackage{microtype}
\usepackage[dvipsnames]{xcolor}
\usepackage[colorlinks=true,linkcolor=MidnightBlue,citecolor=MidnightBlue,urlcolor=MidnightBlue]{hyperref}

\newcommand{\dd}{\mathrm d}
\newcommand{\Tr}{\mathrm{Tr}}
\newcommand{\Diag}{\mathrm{Diag}}

\begin{document}

\title{Reference-Density Hartree Screening for Gausslet Hamiltonians}

\author{Steven R. White}
\affiliation{Department of Physics and Astronomy, University of California, Irvine, Irvine, CA 92697, USA}

\date{September 1, 2026}

\begin{abstract}
Gausslets are among the few electronic-structure bases that permit the
four-index electron--electron interaction to be replaced by
an accurate two-index integral diagonal approximation (IDA).  Near a
many-electron nucleus, however, the nuclear attraction and core-electron
Hartree field are individually large and substantially cancel.
Treating the first as a full finite-basis matrix while treating the second
with IDA leaves an avoidable imbalance.  We introduce reference-density
Hartree screening: the Hartree field of a chosen reference density is
represented accurately, and IDA is applied only to density fluctuations
about it.
Tests on He, Ne, atomic F, F$_2$, and Cr$_2$ show large reductions in
direct Hartree errors, including transfer of fitted neutral-atom fields to
molecules.  For Cr$_2$, atomic-core screening prevents the spurious HF
collapse found with the unscreened $q=5$ and $q=7$ Hamiltonians, whereas
finite-reference matching without screening does not.  Screening leaves
exchange and residual correlation unchanged.  We therefore also
introduce a low-rank one-particle correction that uses an accurate
conventional Gaussian-basis Hartree--Fock calculation to match either
occupied-space exchange information or the complete occupied Fock vectors.
In F$_2$ and Cr$_2$, $X_{\rm HF}$ reproduces the
finite-reference energy and occupied Fock vectors to numerical precision and
the selected states return after orbital perturbations.  For Cr$_2$, the
corrected $q=5$ basis uses one quarter as many functions as the $q=7$
control while retaining sub-mHa mean-field accuracy.  Screening provides the
physical improvement to the direct field; the state-specific correction then
restores the remaining accuracy of the Gaussian-basis mean-field reference.
\end{abstract}

\maketitle

\section{Introduction and background}
\label{sec:intro}

Electronic-structure calculations replace the continuous spatial dependence
of the electrons by a finite representation.  The choice of representation
controls not only how many functions are needed, but also the form and cost of
the electron--electron interaction.  Atom-centered Gaussian orbitals give a
compact description of molecular wave functions, but their Coulomb integrals
form a four-index tensor.  Real-space grids make local potentials and Coulomb
interactions much simpler, but require many points to resolve the different
length scales near a nucleus and in the outer parts of a molecule.  An
ideal representation would combine the compactness and variational
control of a basis set with the locality and simple interaction of a grid.

Gausslets were introduced for this purpose
\cite{White2017Gausslets,WhiteStoudenmire2019}.  They are smooth, localized,
orthonormal functions that behave approximately like broadened, weighted
delta functions when integrated against smooth functions.  This property allows the
four-index electron--electron interaction to be replaced accurately by a
two-index density--density interaction.  Cartesian, hybrid
gausslet--Gaussian, and nested gausslet bases have been applied to atoms,
molecules, and hydrogen chains
\cite{WhiteStoudenmire2019,QiuWhite2021,WhiteLindsey2023}.  Radial and angular
gausslets provide related atom-centered constructions
\cite{White2026RadialGausslets,White2026AngularGausslets}.

\begin{figure}[t]
\includegraphics[width=\columnwidth]{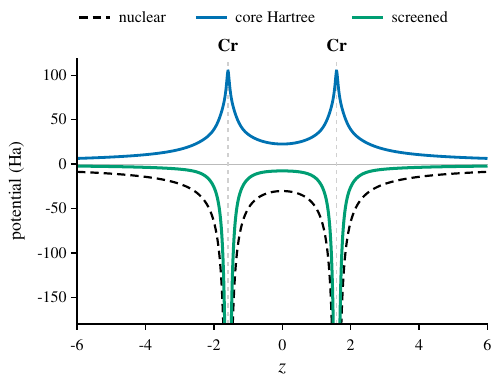}
\caption{\label{fig:screening-field}Reference-density screening along the
Cr$_2$ molecular axis at $R=3.175$ bohr.  Shown are the nuclear attraction
$u_{\rm N}$, the sum $v_{\rm H}^{\rm core}$ of two fitted spherical
18-electron Cr$^{6+}$ Hartree potentials, and their screened sum
$u_{\rm N}+v_{\rm H}^{\rm core}$.}
\end{figure}

Near a many-electron nucleus, the nuclear attraction and core-electron
Hartree repulsion are both large.  Just outside the core they largely cancel,
leaving a much smaller screened potential for the valence electrons.
Figure~\ref{fig:screening-field} shows this along the Cr$_2$ molecular axis.
The cancellation causes no problem if both terms are treated exactly within
the chosen basis.  Once an approximation is introduced, however, it should
be applied after the cancellation.

Standard gausslet Hamiltonians do not do this.  The one-particle nuclear
potential is represented by the full Galerkin matrix
$(U_{\rm N})_{ij}=\langle\chi_i|u_{\rm N}|\chi_j\rangle$.  By contrast,
electron--electron interactions use the integral diagonal approximation
(IDA), of the form
$V_{ij}\hat n_i\hat n_j$, with $V_{ij}$ evaluated as Coulomb
integrals.  Just outside the core of a large atom, the nuclear attraction
therefore contains substantial off-diagonal matrix elements, which cancel
only imperfectly against the diagonal core-Hartree repulsion.
The resulting error can dominate the physical screened field.
For Cr$_2$, this is not a small perturbation: UHF relaxation with the
unscreened $q=5$ and $q=7$ Hamiltonians leads to spurious
residual-Gaussian-dominated states 23.44 and 0.583 Ha, respectively, below
the high-accuracy Gaussian-basis HF reference.  Matching the selected
occupied Fock vectors without screening still leaves false basins at least 20.67 and
0.482 Ha below that reference.  The screened Hamiltonians remain in the
intended orbital basin.

This likely explains why previous calculations with heavier atoms required
unusually high local gausslet order even when the one-particle basis itself
was already adequate \cite{WhiteLindsey2023}.

Reference-density Hartree screening restores the proper order of operations.
We start from the standard exact mean-field rearrangement
$\hat n_i=\langle\hat n_i\rangle_0+\delta\hat n_i$, where
$\delta\hat n_i\equiv
\hat n_i-\langle\hat n_i\rangle_0$.  After this rearrangement, the
diagonal approximation is applied only to
$V_{ij}\delta\hat n_i\delta\hat n_j$.  The Hartree potential and
self-energy of the reference density are instead evaluated accurately.
The nuclear attraction remains Galerkin, so the cancellation between the
nucleus and the reference Hartree field occurs at the one-particle level.
To make the Galerkin reference calculation feasible, we use spherically
symmetric reference densities or sums of them.  For an atomic calculation,
the reference may be a complete closed-shell atom, an atomic core, or a
spherical ensemble density for an open-shell atom.  For a molecular
calculation, the screening density is a sum of such spherical atomic
densities.

Reference atomic densities are also used to reorganize electrostatics in
real-space density-functional calculations.  SIESTA combines each local ionic
pseudopotential with the Hartree potential generated by a spherical atomic
valence density, producing a short-ranged neutral-atom potential
\cite{Soler2002SIESTA}.  Its real-space grid then handles the much smaller
difference between the molecular density and the sum of atomic densities,
greatly reducing finite-grid errors.  FHI-aims makes an analogous all-electron
split \cite{Blum2009FHIaims}.  Spherical free-atom densities and their
electrostatic potentials are obtained accurately from radial calculations,
and only the difference density is treated through atom-centered multipoles.
In both methods the large attractive and Hartree fields are combined before
the smaller remainder is represented numerically.  There the split evaluates
a self-consistent electrostatic potential; here it reorganizes a fixed
many-electron Hamiltonian before IDA is applied, so that only the interaction
between density fluctuations is treated diagonally.

The purpose of Hartree screening also differs from several familiar ways of
reducing core or Coulomb costs.  Effective-core potentials and frozen-core
approximations remove or hold fixed selected core degrees of freedom
\cite{Hamann1979Pseudopotentials,Yu2021FrozenCore}.  Hartree screening keeps all
electrons active.  Our compact Gaussian representations
of atomic densities and potentials are related to density fitting
\cite{Whitten1973DensityFit,Dunlap1979DensityFit}, but they represent a
transferable one-center reference field rather than the molecular four-index
interaction.

Hartree screening supplies the main physical correction.  Remaining exchange
errors can be corrected through single-particle terms.  An accurate
conventional Gaussian-basis Hartree--Fock calculation supplies the target
exchange contribution for the occupied orbitals or, if desired, their
complete Fock vectors.
We impose the corresponding difference with a low-rank Hermitian one-particle
operator and a separate energy constant.  In the full form, this removes all
remaining one-particle error for the selected determinant relative to the
Gaussian-basis calculation.  This correction is related in
purpose to the Qiu--White reference correction and to adaptively compressed
exchange \cite{QiuWhite2021,Lin2016ACE}.  It is less general than Hartree
screening because it is tied to a selected occupied space, but within that
space it gives a more accurate mean-field Hamiltonian.  The Hartree--Fock
target is distinct from the screening density and may be a nonspherical
molecular determinant; correcting the Fock vectors of its occupied orbitals
does not require a complete molecular screening field.

Section~\ref{sec:screening-method} develops reference-density Hartree
screening and the fitted atomic fields used in the calculations, while
Sec.~\ref{sec:hf-completion} introduces the exchange and complete
Hartree--Fock corrections.  We test screening in He, Ne, atomic F, F$_2$,
and Cr$_2$, proceeding from an
isolated one-center mechanism to relaxed molecular calculations and a
heavy-atom example.  F$_2$ and Cr$_2$ also test the accuracy and limits of
correcting either exchange or the complete occupied Fock vectors.

\section{Reference-density Hartree screening}
\label{sec:screening-method}

For the valence electrons of a high-$Z$ atom, the physical mean field is the
screened sum of the nuclear attraction and core-electron Hartree repulsion,
not either large term separately.  The two terms grow with $Z$ and largely
cancel.  In the standard gausslet Hamiltonian, however, the nuclear term is a
full Galerkin matrix while the compensating Hartree field is generated by the
approximate diagonal interaction.  Even a small fractional error in the
Hartree field can then leave a sizable spurious unscreened potential.

This suggests combining the nuclear and reference-Hartree fields before
approximating the remaining interaction.  The required reorganization is the
standard exact mean-field rearrangement.  Choose a reference density
$\rho_0$ and write the density operator as
\begin{equation}
 \hat\rho=\rho_0+\delta\hat\rho .
 \label{eq:density-split}
\end{equation}
For the Coulomb bilinear form
$(f|g)=\iint f(\mathbf r)g(\mathbf r')/
|\mathbf r-\mathbf r'|\,\dd\mathbf r\,\dd\mathbf r'$, the electron--electron
operator is $\hat V_{ee}=\tfrac12:\!(\hat\rho|\hat\rho)\!:$, where the
colons omit the self-contraction of an electron with itself.  The exact
rearrangement is
\begin{equation}
 \frac12:\!(\hat\rho|\hat\rho)\!:
 =(\hat\rho|\rho_0)
 +\frac12:\!(\delta\hat\rho|\delta\hat\rho)\!:
 -\frac12(\rho_0|\rho_0).
 \label{eq:mean-field-identity}
\end{equation}
No approximation has been made in Eq.~\eqref{eq:mean-field-identity}.  We
evaluate the reference-density field and self-energy accurately and apply the
diagonal approximation only to the quadratic fluctuation term.  In the final
basis, let $q_p^0=\langle\hat n_p\rangle_0$ and
$\delta\hat n_p=\hat n_p-q_p^0$.  The screened electron--electron
interaction is then
\begin{equation}
 \begin{aligned}
 \hat V_{ee}^{\rm scr}={}&
 \frac12\sum_{pq}V_{pq}
 (\hat n_p-q_p^0)(\hat n_q-q_q^0)
 -\frac12\sum_pV_{pp}\hat n_p
 \\
 &+\sum_{pq\sigma}(J_0)_{pq}
 a_{p\sigma}^\dagger a_{q\sigma}-E_{\rm H}^0.
 \end{aligned}
 \label{eq:screened-interaction}
\end{equation}
Here $J_0$ and $E_{\rm H}^0$ are the accurately represented Hartree matrix
and self-energy of the reference density.  The diagonal two-particle matrix
$V_{pq}$ therefore acts on density fluctuations about that reference, while
the reference Hartree field itself is placed in the one-particle Hamiltonian.

\subsection{The diagonal interaction and its reference field}

Let $\{\chi_p\}$ be the final real orthonormal working basis, consisting of
the contracted gausslets together with the retained residual Gaussian
directions.  Kinetic energy and nuclear attraction are evaluated as Galerkin
matrices.  We denote the physical one-electron operator and its matrix in the
working basis by
\begin{align}
 u_{\rm N}(\mathbf r)&=-\sum_A\frac{Z_A}{|\mathbf r-\mathbf R_A|},
 \nonumber\\
 \hat h&=-\frac12\nabla^2+u_{\rm N}(\mathbf r),
 \nonumber\\
 h_{pq}&=\langle\chi_p|\hat h|\chi_q\rangle
 =T_{pq}+(U_{\rm N})_{pq},
 \label{eq:physical-h1}
\end{align}
where $(U_{\rm N})_{pq}=\langle\chi_p|u_{\rm N}|\chi_q\rangle$.
The full nuclear matrix $U_{\rm N}$ is retained throughout.
The nuclear--nuclear repulsion is omitted from the operator equations below,
since it is unchanged by screening.

For the gausslet--gausslet electron--electron interaction we use the integral
diagonal approximation (IDA) introduced with the original gausslets
\cite{White2017Gausslets}.  It has a two-index density--density form, but its
matrix is defined by Coulomb integrals over the individual gausslets.  If
$G_i(\mathbf r)$ is a gausslet and
$w_i=\int G_i(\mathbf r)\,\dd\mathbf r$, then
\begin{equation}
 V_{ik}=\frac{1}{w_iw_k}
 \int\!\!\int G_i(\mathbf r)
 \frac{G_k(\mathbf r')}{|\mathbf r-\mathbf r'|}
 \,\dd\mathbf r\,\dd\mathbf r'.
 \label{eq:ida-integral}
\end{equation}
Thus IDA averages the Coulomb kernel over the finite extent of both gausslets
rather than evaluating it only at their centers.  Writing the resulting
two-index interaction matrix as $V$, the spin-summed interaction is
\begin{equation}
 \hat V_{ee}
 =\frac12\hat{\bm n}^{T}V\hat{\bm n}
 -\frac12\sum_p V_{pp}\hat n_p,
 \label{eq:ida-operator}
\end{equation}
where $\hat n_p=\sum_\sigma
a_{p\sigma}^\dagger a_{p\sigma}$.  The last term removes same-orbital
self-interaction.  PQS contractions and residual Gaussian terms are
incorporated into the final two-index matrix $V$ using the interaction
constructions described in Refs.~\cite{WhitePQS2026,QiuWhite2021}.  The
screening derivation below depends only on this final matrix.  Together with
the physical one-electron matrix $h$, it defines the ordinary unscreened
gausslet Hamiltonian,
\begin{equation}
 \hat H
 =\sum_{pq\sigma}h_{pq}a_{p\sigma}^\dagger a_{q\sigma}
 +\hat V_{ee}.
 \label{eq:H-unscreened}
\end{equation}

The screening charge is specified by reference spatial orbitals $\phi_a$ and
their spin-summed occupations $f_a$.  For a closed-shell atom or atomic core,
the occupied orbitals have $f_a=2$.  The open-shell ensemble described below
can have fractional occupations.  In either case, the occupations are chosen
so that the density of each atom is spherical.  The gausslet sector alone does
not generally span the reference orbitals exactly.  We therefore choose the
Gaussian supplement to contain their missing components: after projecting the
Gaussian basis used to calculate $\phi_a$ against the gausslet sector, we
retain the linearly independent residual directions needed to recover every
$\phi_a$.  The combined working basis $\{\chi_p\}$ therefore contains the
reference orbitals precisely.  Their expansions in this basis define the
coefficient matrix $C$ and the spin-summed one-particle density matrix $P^0$:
\begin{equation}
 \begin{aligned}
 \phi_a(\mathbf r)&=\sum_p C_{pa}\chi_p(\mathbf r),\\
 P^0_{pq}&=\sum_a f_a C_{pa}C_{qa},\\
 \rho_0(\mathbf r)&=\sum_a f_a|\phi_a(\mathbf r)|^2
 =\sum_{pq}P^0_{pq}\chi_p(\mathbf r)\chi_q(\mathbf r),\\
 q_p^0&=P^0_{pp}.
 \end{aligned}
 \label{eq:reference-density}
\end{equation}
The entries $q_p^0$ form the vector $\bm q^0$, and
$\delta\hat{\bm n}=\hat{\bm n}-\bm q^0$.
The accurately represented reference Hartree field and self-energy are
\begin{align}
 (J_0)_{pq}&=(\chi_p\chi_q|\rho_0),
 &E_{\rm H}^0&=\frac12(\rho_0|\rho_0).
 \label{eq:reference-field}
\end{align}
They combine with the nuclear attraction to form the screened potential
matrix
\begin{equation}
 W_0=U_{\rm N}+J_0.
 \label{eq:screened-field}
\end{equation}
This full matrix sum contains the large nuclear--Hartree cancellation that
motivates the method, including its short-range off-diagonal terms.

When $J_0$ and $E_{\rm H}^0$ are evaluated directly from the same represented
density, they obey
\begin{equation}
 E_{\rm H}^0=\frac12\Tr(P^0J_0).
 \label{eq:exact-field-closure}
\end{equation}
For the fitted atomic fields used in our molecular calculations, the density
fit determines $E_{\rm H}^0$, and a separate fit of its Hartree potential
determines $J_0$.

The unscreened Hamiltonian represents the same reference charge $\bm q^0$ by
the approximate diagonal Hartree field and self-energy
\begin{align}
 J_0^{\rm IDA}&=\Diag(V\bm q^0),
 &E_{\rm H}^{\rm IDA,0}&=\frac12(\bm q^0)^T V\bm q^0.
 \label{eq:ida-reference}
\end{align}
Screening replaces these baseline reference terms by $J_0$ and
$E_{\rm H}^0$.  Relative to the unscreened Hamiltonian, the required
corrections are
\begin{align}
 \Delta J_0&=J_0-\Diag(V\bm q^0),
 \label{eq:delta-J0}\\
 C_{\rm H}&=\frac12(\bm q^0)^T V\bm q^0-E_{\rm H}^0.
 \label{eq:CH}
\end{align}
Although screening changes how the electron--electron interaction is
approximated, in the Hamiltonian it changes only the one-particle operator by
$\Delta J_0$ and adds the constant $C_{\rm H}$.  The
screened one-electron matrix is therefore
\begin{equation}
 h_{\rm scr}=h+\Delta J_0,
 \label{eq:hscr}
\end{equation}
and the screened many-electron Hamiltonian is
\begin{equation}
 \begin{aligned}
 \hat H_{\rm scr}
 &=\hat H
 +\sum_{pq\sigma}(\Delta J_0)_{pq}
 a_{p\sigma}^\dagger a_{q\sigma}+C_{\rm H}\\
 &=\sum_{pq\sigma}(h_{\rm scr})_{pq}
 a_{p\sigma}^\dagger a_{q\sigma}
 +\hat V_{ee}+C_{\rm H}.
 \end{aligned}
 \label{eq:Hscr-correction}
\end{equation}
Equivalently, collecting the nuclear and reference-Hartree fields gives
\begin{equation}
 \begin{aligned}
 \hat H_{\rm scr}={}&\sum_{pq\sigma}
 \bigl[T+W_0\bigr]_{pq}a_{p\sigma}^\dagger a_{q\sigma}
 +\frac12\delta\hat{\bm n}^{T}V\delta\hat{\bm n}\\
 &-\frac12\sum_pV_{pp}\hat n_p-E_{\rm H}^0.
 \end{aligned}
 \label{eq:Hscr-centered}
\end{equation}
Equation~\eqref{eq:Hscr-correction} adds the two correction terms to the
unscreened Hamiltonian.  Rewriting the same Hamiltonian in terms of
$\delta\hat{\bm n}$ gives Eq.~\eqref{eq:Hscr-centered}: nuclear attraction
and reference Hartree repulsion cancel within the accurate one-particle field
$W_0$, and the approximate two-particle interaction acts on fluctuations
about the reference density.

At the Hartree level, this rearrangement also gives a useful statement about
the remaining direct-energy error.  Let $\mathcal E_J[\rho]$ denote the
Galerkin Hartree energy of a represented density and
$\mathcal E_J^{\rm IDA}[\rho]$ its IDA counterpart.  For
$\rho=\rho_0+\delta\rho$, and with $J_0$ and $E_{\rm H}^0$ evaluated exactly
for $\rho_0$, the screened approximation obeys
\begin{equation}
 \mathcal E_{J,{\rm scr}}^{\rm IDA}[\rho]
 -\mathcal E_J[\rho]
 =\mathcal E_J^{\rm IDA}[\delta\rho]
 -\mathcal E_J[\delta\rho].
 \label{eq:quadratic-direct-error}
\end{equation}
The constant term and the term linear in $\delta\rho$ are supplied by
$E_{\rm H}^0$ and $J_0$.  The remaining direct error is therefore second
order in the difference between the actual and reference densities.  This is
why a sum of spherical atomic fields can remain useful in a molecule even
though it is not the molecular Hartree field.  For fitted fields, the same
statement holds to fitting accuracy.

At that density, the direct Fock field is $J_0$.  The corresponding direct
energy is $\Tr(P^0J_0)-E_{\rm H}^0$, which equals $E_{\rm H}^0$ when
Eq.~\eqref{eq:exact-field-closure} holds.  For the separately fitted fields,
the difference is the fitting inconsistency
$\Tr(P^0J_0)-2E_{\rm H}^0$.

At a reference determinant, each
$\langle\delta\hat n_p\rangle_0$ vanishes, but the products
$\langle\delta\hat n_p\delta\hat n_q\rangle_0$ generally do not.  The
quadratic fluctuation term in Eq.~\eqref{eq:Hscr-centered} therefore remains
nonzero.

\subsection{Atomic reference fields}

For a closed-shell atom, the natural all-electron reference is its spherical
Hartree--Fock density.  A core-only reference can be used when the purpose is
to remove only the most compact part of the field.  For an open-shell atom, a
single determinant generally chooses an orientation.  We instead average the
fractional occupation of a partially filled shell over its magnetic
components.  This gives a neutral spherical density that can be translated
and superposed without assigning an orientation to each atom in a molecule.

All molecular screening calculations below use an atom-centered sum
\begin{equation}
 \rho_0(\mathbf r)=\sum_A\rho_A(|\mathbf r-\mathbf R_A|)
 \label{eq:atomic-density-sum}
\end{equation}
of such spherical densities, whose Hartree field is the sum of their
translated radial potentials.

Direct evaluation of the complete reference Hartree matrix is too expensive
for large atoms or molecules.  We fit each spherical atomic density in the
form
$\rho_A(r)\simeq\sum_\mu c_\mu(\alpha_\mu/\pi)^{3/2}
e^{-\alpha_\mu r^2}$, using logarithmically spaced widths and enforcing the
total charge.  Its radial Hartree potential is fitted separately as
$v_A(r)\simeq\sum_\nu d_\nu e^{-\beta_\nu r^2}$; the calculations below use
33 potential terms.  The density fit determines $E_{\rm H}^0$, while the
potential fit determines $J_0$.  The coefficients and exponents are retained
with the numerical records.  A few small-system tests evaluate the reference
field directly, but practical calculations use the fitted form.  The same
fitted density and potential can be translated to each nucleus and reused at
different geometries.

\section{Finite-reference Hartree--Fock matching}
\label{sec:hf-completion}

Reference-density screening fixes the direct Hartree field at the chosen
density, but it leaves the exchange field of the approximate interaction
unchanged.  An accurate conventional Gaussian-basis Hartree--Fock calculation
can provide more accurate information for these remaining one-particle terms.
With a low-rank one-particle correction, we can match either its exchange
field alone or its complete occupied Fock vectors.  In the latter case, the selected
determinant becomes stationary for the reference-matched Hamiltonian.  This
determinant need not have the spherical atomic form used to construct the
screening density.

For clarity, we first write the construction for a restricted determinant,
with all matrices represented in a common orthonormal working basis.  Let the
columns of $Q$ be the $n_o$ occupied spatial orbitals of the selected
reference determinant, with $Q^\dagger Q=I_{n_o}$, and let
$Y=F_{\rm ref}Q$ be their reference Fock vectors.  We write $\Phi_Q$ for the
determinant in which these spatial orbitals are doubly occupied.

Let $F$ be the Fock matrix of the screened gausslet Hamiltonian, before the
finite-reference correction, evaluated at $\Phi_Q$.  We define the projector
onto its occupied space and the complementary projector by
\begin{equation}
 \Pi=QQ^\dagger,
 \qquad
 \bar\Pi=I-\Pi.
 \label{eq:occupied-projectors}
\end{equation}
We want the corrected Fock matrix to equal $F_{\rm ref}$ whenever either side
lies in the occupied space, while leaving the
orthogonal--orthogonal block of $F$ unchanged.  These requirements are
expressed directly by
\begin{equation}
 F_{\rm corr}=F_{\rm ref}
 -\bar\Pi(F_{\rm ref}-F)\bar\Pi.
 \label{eq:corrected-fock-projector}
\end{equation}
Only blocks touching the occupied space enter this equation.  Thus
$F_{\rm ref}$ may be any Hermitian extension of the prescribed vectors
$Y=F_{\rm ref}Q$; its orthogonal--orthogonal block cancels from the result.
Writing $\Delta F=F_{\rm ref}-F$ and expanding $\bar\Pi=I-\Pi$ gives
\begin{align}
 F_{\rm corr}&=F+X_{\rm HF},\notag\\
 X_{\rm HF}&=\Pi\Delta F+\Delta F\Pi-\Pi\Delta F\Pi.
 \label{eq:XHF-projector}
\end{align}

Only the occupied columns of $\Delta F$ enter this result.  Defining
\begin{equation}
 \Delta Y=Y-FQ=\Delta FQ,
 \label{eq:delta-Y}
\end{equation}
Eq.~\eqref{eq:XHF-projector} becomes
\begin{equation}
 X_{\rm HF}=\Delta YQ^\dagger+Q\Delta Y^\dagger
 -Q(Q^\dagger\Delta Y)Q^\dagger.
 \label{eq:minimal-lift}
\end{equation}
Equation~\eqref{eq:minimal-lift} obeys
\begin{equation}
 (F+X_{\rm HF})Q=Y.
 \label{eq:fock-orbital-match}
\end{equation}
The rank of $X_{\rm HF}$ is at most $2n_o$, since its matrix elements between
two orbitals both orthogonal to the occupied space are zero.

A more limited correction can be used when only the occupied--occupied Fock
block is wanted:
\begin{equation}
 X_{oo}=\Pi\Delta F\Pi
 =Q(Q^\dagger\Delta Y)Q^\dagger.
 \label{eq:occupied-block-correction}
\end{equation}
This form reproduces the reference Fock matrix within the occupied space, but
does not correct its couplings to the orthogonal space and therefore does not
make the reference determinant stationary.

The Gaussian-basis Hartree--Fock calculation begins in its own orthonormal
one-particle space.  We project its occupied orbitals $Q_{\rm ref}$ and Fock
vectors $Y_{\rm ref}=F_{\rm ref}Q_{\rm ref}$ into the gausslet space and retain
the missing components as residual Gaussian supplements.  When all of these
directions are retained, the combined basis contains the reference orbitals
and Fock vectors precisely; their coefficient matrices are the $Q$ and $Y$
used above.  A minimal supplement for this purpose has at most $2n_o$
directions, even when the reference functions contain many angular-momentum
components.  The calculations below generally retain a larger residual
Gaussian sector because the same working basis is also used for correlated
calculations.  For Cr$_2$, by contrast, we deliberately restrict the residual
supplement and apply the construction to captured projections of the
reference orbitals and Fock vectors, as described below.  In either case, only
the occupied orbitals and their Fock vectors need to be transferred; the
complete reference Fock matrix in the working basis is never constructed.

For an unrestricted reference, the same construction is applied separately
to the $\alpha$ and $\beta$ occupied orbitals, producing spin-dependent
matrices $X^\alpha$ and $X^\beta$.  The Cr$_2$ calculations below use this
spin-resolved form.

Returning to the restricted case, the full-Fock matching correction
$X_{\rm HF}$ uses the complete reference Fock vectors $Y$.  It is added to the screened
one-electron matrix,
\begin{equation}
 h_{\rm HF}=h_{\rm scr}+X_{\rm HF},
 \label{eq:hHF}
\end{equation}
while the two-particle operator $\hat V_{ee}$ is left unchanged.
The corresponding second-quantized one-particle correction is
\begin{equation}
 \hat X_{\rm HF}=\sum_{pq\sigma}(X_{\rm HF})_{pq}
 a_{p\sigma}^\dagger a_{q\sigma}.
 \label{eq:XHF-operator}
\end{equation}
To also reproduce the electronic energy of the finite Hartree--Fock
reference, we add the constant
\begin{equation}
 C_{\rm HF}=E_{\rm HF}^{\rm ref}-\langle\Phi_Q|
 \hat H_{\rm scr}+\hat X_{\rm HF}|\Phi_Q\rangle .
 \label{eq:completion-scalar}
\end{equation}
Molecular total energies include the nuclear--nuclear repulsion separately.
The completed many-electron Hamiltonian is therefore
\begin{equation}
 \hat H_{\rm HF}=\hat H_{\rm scr}+\hat X_{\rm HF}+C_{\rm HF}.
 \label{eq:Hhf}
\end{equation}
With $X_{\rm HF}$ and $C_{\rm HF}$, the selected reference energy, all occupied
Fock vectors, and all occupied-to-orthogonal couplings agree with that finite
Gaussian-basis Hartree--Fock calculation.  When the conventional calculation
is well converged in a large Gaussian basis, it provides an accurate
mean-field anchor and a stationary determinant.

To isolate the exchange error left unchanged by Hartree screening, we also
form an exchange-only correction.  This is primarily a diagnostic because it
changes only the exchange component.  The exchange contribution to the Fock
matrix is $-K$.  The difference between the reference and model exchange
vectors is most directly expressed without constructing a complete reference
exchange matrix in the working basis.  Let $Y_{\rm x}^{\rm ref}$ be the
reference exchange Fock vectors, $-K_{\rm ref}Q_{\rm ref}$, represented in the
working basis.  Since the model exchange vectors are $-KQ$, their difference
is
\begin{equation}
 \Delta Y_{\rm x}=Y_{\rm x}^{\rm ref}+KQ.
 \label{eq:exchange-orbital-residual}
\end{equation}
Here $Q^\dagger\Delta Y_{\rm x}$ is Hermitian because it is the difference between
the reference and model exchange matrices within the occupied space.  Using
$\Delta Y_{\rm x}$ in Eq.~\eqref{eq:minimal-lift} therefore defines $X_{\rm x}$,
which is added to $h_{\rm scr}$ in place of $X_{\rm HF}$.  If
$\Delta E_{\rm x}=E_{\rm x}^{\rm ref}-E_{\rm x}$, the accompanying constant is
\begin{equation}
 C_{\rm x}=-\Delta E_{\rm x}.
 \label{eq:exchange-scalar}
\end{equation}
The one-particle exchange field contributes twice its desired energy change
at the reference determinant, and $C_{\rm x}$ removes the extra copy.
Together, $X_{\rm x}$ and $C_{\rm x}$ reproduce the reference exchange energy
and the reference exchange Fock vectors for every occupied orbital.  The
direct field remains unchanged.

The scope of this construction can also be seen by normal ordering the
difference $\Delta\hat V=\hat V_{\rm ref}-\hat V_{\rm model}$ between a
reference and a model two-particle interaction about the reference
determinant:
\begin{equation}
 \Delta\hat V=\Delta E^0
 +\sum_{pq}\Delta f^0_{pq}\{a_p^\dagger a_q\}_0
 +\Delta\hat V_{\rm res}^{(2)}.
 \label{eq:normal-decomposition}
\end{equation}
Here $p$ and $q$ label spin orbitals,
$\Delta E^0=\langle\Phi_Q|\Delta\hat V|\Phi_Q\rangle$, and
$\Delta f^0$ is the Fock-matrix difference obtained by contracting
$\Delta\hat V$ once with the reference density matrix.  The braces denote
normal ordering with respect to $\Phi_Q$; for example,
$\{a_p^\dagger a_q\}_0=a_p^\dagger a_q-
\langle\Phi_Q|a_p^\dagger a_q|\Phi_Q\rangle$, and all contractions are
subtracted from higher products.  The constant and one-particle terms fix the
reference energy and all couplings from $\Phi_Q$ to single excitations.  The
matrix elements of the residual two-particle operator from $\Phi_Q$ begin
with double excitations.
$X_{\rm x}$ or $X_{\rm HF}$ supplies the chosen one-particle information,
but neither reconstructs this residual interaction or prescribes a correction
for a different determinant or a correlated state.

This construction is related to the stationary-Fock correction of Qiu and
White and to adaptively compressed exchange
\cite{QiuWhite2021,Lin2016ACE}, but the reference calculation plays a more
central role here.  We take an accurate conventional Gaussian-basis
Hartree--Fock calculation as the desired mean-field result and choose
$X_{\rm HF}$ so that the gausslet Hamiltonian reproduces its energy and
occupied Fock vectors.

In principle, even a very small gausslet basis, for example one with $q<5$,
could be corrected in this way to reproduce a selected Hartree--Fock
reference.  Such enforced agreement would not by itself make the basis
robust.  The direct screening correction is valuable because it incorporates
an important physical principle: the cancellation between the nuclear
attraction and the Hartree field of the reference density.  Once this large
physical error is removed from the approximate Hamiltonian, its energies and
orbitals improve without being forced to match a particular calculation.
The subsequent $X_{\rm HF}$ correction is more pragmatic.  It uses simple
one-particle terms to make the gausslet Hamiltonian reproduce the remaining
one-particle information from the more accurate reference calculation; its
accuracy comes from that calculation rather than from additional physical
content in the correction.
One practical check on robustness is therefore to rerun Hartree--Fock after
both corrections have been included, starting from the reference orbitals and
from perturbed orbitals, and determine whether the optimized state remains
near the reference or changes substantially.

This approach separates the accuracy of the Hartree--Fock reference from the
basis size needed for the subsequent correlated calculation.  A moderate
gausslet basis can retain the Gaussian-basis mean-field accuracy, while its
much larger set of localized orbitals remains available to describe
correlation.  We expect this additional space to improve correlated
calculations without first increasing $q$ merely to remove mean-field errors.
The correction does not, however, determine the remaining two-particle
interaction, and the orbital-return check establishes only local restricted
robustness, not a global or unrestricted stability result.

\section{Numerical tests}
\label{sec:results}

All calculations reported here use the recently introduced projected
$q$-shell (PQS) form of nested Cartesian gausslets \cite{WhitePQS2026},
augmented where needed by residual Gaussian functions \cite{QiuWhite2021}.
Here and below, a residual Gaussian function is a conventional Gaussian
linear combination orthogonalized against and appended to the PQS basis.
The gausslet basis construction and Hamiltonian generation were performed
with the open-source \texttt{GaussletBases.jl} software library at
\url{https://github.com/srwhite59/GaussletBases.jl}.  All physical
one-particle operators are Galerkin matrices.  Independent Gaussian reference
determinants and their one- and two-electron quantities were generated with
PySCF \cite{Sun2018PySCF}.  At a selected reference determinant $\Phi_0$,
we define the component errors
\begin{equation}
 \Delta E_A=E_A^{\rm model}[\Phi_0]-E_A^{\rm ref}[\Phi_0],
 \qquad A=J,K .
 \label{eq:component-errors}
\end{equation}
Here $E_J$ is the direct Hartree energy and $E_K$ is the signed exchange
energy, which is negative in the usual convention.
If the physical one-particle expectation also differs, we denote its error by
$\Delta E_h$; at the fixed determinant,
$\Delta E_{\rm HF}=\Delta E_h+\Delta E_J+\Delta E_K$.
The screening potential and its scalar are included in $E_J$, since both are
obtained from the reference-density rearrangement of the direct interaction.
The finite-reference correction $X_{\rm HF}$ completes the total Fock field
rather than changing a uniquely defined direct or exchange term.  We therefore
do not assign separate $\Delta E_J$ or $\Delta E_K$ values to rows containing
this correction.  Some tests hold the orbitals fixed
to isolate the component errors, while others reoptimize them.  In the tables,
an arrow gives the fixed-reference error followed by the outcome of an
independent Hartree--Fock optimization.

\begin{table*}[t]
\caption{\label{tab:he}Component and energy errors for He, in mHa.  The
direct and exchange errors and the RHF arrows use the finite cc-pV6Z
($l\leq1$) reference; this reference is $0.007$ mHa above the
nonrelativistic continuum RHF energy \cite{Cinal2020}.  Each arrow gives the
error at the embedded reference determinant followed by the error after
independent RHF optimization.  The correlation-energy error is relative to
the continuum value \cite{Nakashima2007He}, with
$E_{\rm corr}=E_{\rm gs}-E_{\rm RHF}$ for each finite Hamiltonian.  $N$
includes the 21 residual Gaussian functions.}
\begin{ruledtabular}
\begin{tabular}{rrlrrcr}
$q$ & $N$ & Hamiltonian & $\Delta E_J$ & $\Delta E_K$
& $\Delta E_{\rm RHF}$, fixed $\rightarrow$ relaxed
& $\Delta E_{\rm corr}$\\
\hline
4 & 482  & unscreened & $-4.18$ & $2.09$ & $-2.09\rightarrow-2.14$ & $0.735$\\
  &      & screened   & $0.000$ & $2.09$ & $2.09\rightarrow2.04$ & $0.706$\\
  &      & screened+$X_{\rm HF}$ & --- & --- & $0\rightarrow0$ & $0.752$\\
5 & 832  & unscreened & $0.174$ & $-0.087$ & $0.087\rightarrow0.086$ & $0.223$\\
  &      & screened   & $0.000$ & $-0.087$ & $-0.087\rightarrow-0.088$ & $0.225$\\
  &      & screened+$X_{\rm HF}$ & --- & --- & $0\rightarrow0$ & $0.223$\\
6 & 1428 & unscreened & $-0.0492$ & $0.0246$ & $-0.0246\rightarrow-0.0257$ & $0.100$\\
  &      & screened   & $0.000$ & $0.0246$ & $0.0246\rightarrow0.0232$ & $0.100$\\
  &      & screened+$X_{\rm HF}$ & --- & --- & $0\rightarrow0$ & $0.101$\\
\end{tabular}
\end{ruledtabular}
\end{table*}

\begin{table*}[t]
\caption{\label{tab:ne}Hartree screening and reference-HF correction for
Ne.  Conventions follow Table~\ref{tab:he}.  The reference is the
finite cc-pV6Z ($l\leq1$) RHF result, $-128.547061$ Ha; the numerical HF limit is
$-128.547098$ Ha \cite{Cinal2020}, only $0.037$ mHa lower.}
\begin{ruledtabular}
\begin{tabular}{rrlrrc}
$q$ & $N$ & Hamiltonian & $\Delta E_J$ & $\Delta E_K$
& $\Delta E_{\rm RHF}$, fixed $\rightarrow$ relaxed\\
\hline
5 & 1326 & unscreened
  & $-6.39$ & $2.46$ & $-3.94\rightarrow-5.09$\\
  &      & $1s^2$-core screened
  & $-4.95$ & $2.46$ & $-2.49\rightarrow-3.04$\\
  &      & neutral-atom screened
  & $0.001$ & $2.46$ & $2.46\rightarrow2.39$\\
  &      & neutral-atom screened $+X_{\rm HF}$
  & --- & --- & $0\rightarrow0$\\
7 & 3420 & unscreened
  & $-0.130$ & $0.041$ & $-0.089\rightarrow-0.109$\\
  &      & $1s^2$-core screened
  & $-0.092$ & $0.041$ & $-0.051\rightarrow-0.068$\\
  &      & neutral-atom screened
  & $0.001$ & $0.041$ & $0.042\rightarrow0.029$\\
  &      & neutral-atom screened $+X_{\rm HF}$
  & --- & --- & $0\rightarrow0$\\
\end{tabular}
\end{ruledtabular}
\end{table*}

\begin{table*}[t]
\caption{\label{tab:f2-completion}Screening and finite-reference
Hartree--Fock matching for restricted F$_2$ at $R=2.668$ bohr, using $q=5$
and $N=2553$.  Errors are in mHa.  Other conventions follow the preceding
tables, with the retained rank-250 cc-pV5Z RHF result as the reference.}
\begin{ruledtabular}
\begin{tabular}{lrrc}
Hamiltonian & $\Delta E_J$ & $\Delta E_K$
& $\Delta E_{\rm RHF}$, fixed $\rightarrow$ relaxed\\
\hline
unscreened & $-52.36$ & $15.87$ & $-36.49\rightarrow-46.51$\\
core screened & $-37.39$ & $15.87$ & $-21.52\rightarrow-26.20$\\
neutral-atom screened & $0.865$ & $15.87$ & $16.73\rightarrow15.82$\\
neutral-atom screened+$X_{\rm HF}$ & --- & --- & $0\rightarrow0$\\
\end{tabular}
\end{ruledtabular}
\end{table*}

\subsection{One-center mechanism and its limits: He}

Helium is a useful one-center calibration, but it is an unusually unfavorable
system on which to judge the practical value of Hartree screening.  It has
only two electrons, both in the same spatial orbital, and has no separate
core and valence densities.  Moreover, the total-density Hartree term contains
a same-orbital self-interaction that exchange cancels.  If
$J=(\phi\phi|\phi\phi)$ for the occupied orbital, the total-density Hartree
energy is $2J$ and the exchange energy is $-J$, leaving the physical
opposite-spin interaction $J$.  Hartree screening improves the direct field
but leaves exchange unchanged, so a small total error may improve or worsen
when this cancellation changes.  We therefore use He mainly to test the
screening construction, the full Hartree--Fock correction, and the origin of
the remaining two-electron error.

We use the standard $q=4$, 5, and 6 PQS constructions, with core spacings
0.20, 0.15, and 0.12 bohr, respectively.  Each is augmented by the complete
21 residual Gaussians from a contracted Cartesian cc-pV6Z basis restricted to
$l\leq 1$, giving final dimensions 482, 832, and 1428 \cite{Dunning1989}.
The same Gaussian calculation supplies the spherical RHF reference density.
Its Hartree field is represented by the 33-term fitted potential described in
Sec.~\ref{sec:screening-method}.  For every Hamiltonian we evaluate the
embedded Gaussian determinant, solve the RHF equations independently, and
obtain the exact ground state of the resulting finite two-electron
Hamiltonian.

Table~\ref{tab:he} separates the errors in the mean-field and correlation
energies.  At $q=4$, the unscreened direct and exchange errors partially
cancel.  Screening reduces the direct error from $-4.18$ mHa to the displayed
precision, leaving the $2.09$ mHa exchange error exposed; the reversal of the
total RHF error is therefore not an overcorrection of the direct term.  The
same pattern persists with much smaller component errors at $q=5$ and 6.
The nearly unchanged correlation-energy errors show that these mean-field
corrections do not repair the limited resolution of the electron--electron
cusp.

For every $q$, the maximum absolute element of
$(F+X_{\rm HF})Q-Y$ is below $7.1\times10^{-14}$ Ha.
Independent RHF calculations starting from the reference, the usual Aufbau
orbitals, and deliberately perturbed occupied spaces all return to the
selected determinant.  Taken together, the three orders separate the roles
of the two corrections: Hartree screening repairs the direct field,
$X_{\rm HF}$ supplies the remaining mean-field information from the
reference calculation, and increasing $q$ improves the subsequent correlated
calculation.

The $q=5$ decomposition confirms the same point.  Within the common Gaussian
orbital space, replacing the full Coulomb interaction by the screened
two-index Hamiltonian with $X_{\rm HF}$ changes the ground-state energy by
only 0.023 mHa.  The accuracy of the correlation energy is instead controlled
primarily by basis resolution and size---here by $q$, and consequently
$N$---not by screening or $X_{\rm HF}$.

\subsection{Closed-shell atomic screening: Ne}

Ne is a more representative closed-shell test of Hartree screening.  Its compact $1s^2$
core is distinct from the $2s^22p^6$ valence shell, while its all-electron
neutral-atom density is spherical.  We can therefore compare screening by the
core alone with screening by all ten electrons, without introducing an
ensemble density.

The RHF ground state of Ne was also studied in the original White--Lindsey
nesting paper \cite{WhiteLindsey2023}.  Those calculations used $n_s=7$, 9,
and 11, with sub-mHa accuracy beginning only at $n_s=9$.  With PQS, $q=7$
reaches this accuracy without screening, while the screened and
reference-corrected $q=5$ Hamiltonian does so with substantially fewer
functions.

At the common nominal order $q=n_s=7$, the two bases are comparable in
size and resolution.  Both contain 3402 functions, including seven residual
$s$ Gaussians from cc-pV6Z, although their coordinate mappings differ
slightly.  The RHF error is 6.50 mHa with White--Lindsey nesting and 1.42 mHa
with PQS.  This improvement matches expectations \cite{WhitePQS2026}.
Adding the 18 residual $p$ Gaussians, as in our current practice for
first-row atoms, gives a 3420-function PQS basis and reduces the absolute
unscreened error relative to the numerical HF limit to 0.073 mHa.  The
analogous $q=5$ basis has 1326 functions and a core
spacing of 0.03 bohr, compared with 0.02 bohr at $q=7$; both bases use all 25
residual $s$ and $p$ Gaussians \cite{Dunning1989}.

The same restricted cc-pV6Z calculation supplies the determinant and
reference charge densities.  We compare no screening, $1s^2$-core screening,
all-electron neutral-atom screening, and neutral-atom screening followed by
$X_{\rm HF}$.  The screened fields use the compact density and potential
fits described in Sec.~\ref{sec:screening-method}.  Table~\ref{tab:ne} gives the total RHF error for the
reference determinant, its direct and exchange contributions, and the error
after orbital relaxation.

Errors of a few tenths of a mHa are already highly accurate for Ne, and
unscreened $q=7$ PQS reaches this level.  Its small total error nevertheless
contains a direct error of $-0.130$ mHa partially canceled by exchange.
Neutral-atom screening reduces the direct error to the displayed precision,
leaving the much smaller exchange error visible.

How small a basis can we use and still have sub-mHa errors?  At $q=5$,
unscreened PQS has errors of several mHa, and core-only screening gives a
useful but incomplete improvement.  With neutral-atom screening and
$X_{\rm HF}$, however, the 1326-function calculation reproduces the finite
cc-pV6Z result, itself within $0.037$ mHa of the numerical HF limit.  Orbital
relaxation changes the energy by less than the precision shown, compared with
1.15 mHa without screening.  Direct screening removes most of this drift;
$X_{\rm HF}$ then supplies the remaining total Fock correction.  The resulting
Hamiltonian is therefore a compact, locally robust HF starting point for
correlated calculations.  Its correlation accuracy, including the effect of
short-range cusp incompleteness, remains to be tested.

\subsection{A spherical open-shell reference: F and
\texorpdfstring{F$_2$}{F2}}

For a closed-shell atom the neutral atomic density is already spherical.
Fluorine instead has an oriented beta-spin $p$ hole.  Averaging this hole
equally over its three orientations gives a spherical nine-electron ensemble
density, a rotationally neutral atomic reference that can be transferred to a
molecule.  We represent its density
and Hartree potential by the Gaussian fits described in
Sec.~\ref{sec:screening-method}.

At the orientation-specific atomic UHF determinant, using a standard $q=5$
PQS basis augmented by residual Gaussians to $N=1420$, the unscreened,
core-screened, and neutral-screened direct errors are $-6.13$, $-4.66$, and
$0.333$ mHa.  The exchange error is unchanged at $1.81$ mHa.  Thus the
spherical neutral density gives the decisive correction to the direct energy,
while core-only screening gives a much smaller improvement.

We next translate the same fitted core and neutral-atom fields, without
refitting, to restricted F$_2$ at the near-equilibrium separation
$R=2.668$ bohr ($1.412$~\AA).  Every F$_2$ calculation uses the standard
$q=5$ PQS basis, augmented by residual Gaussians to a total dimension of
2553.  Its Cartesian cc-pV5Z reference has 252 AOs; removing two
near-linearly-dependent overlap directions gives the retained rank-250 RHF
space used here.
The translated atomic field remains effective even though bonding and
polarization make the molecular density different from a sum of neutral
atoms.

For F$_2$, Hartree screening repairs nearly all of the direct-energy error, but
it leaves exchange unchanged.  We next use the same F$_2$ reference to test
the two one-particle matching corrections.  The nine occupied reference
orbitals are embedded in the 2553-dimensional working basis.  Both
$X_{\rm x}$ and $X_{\rm HF}$ have rank 18.  The first matches the occupied
exchange vectors; the second matches the complete occupied Fock vectors and
anchors the finite-reference RHF energy.

Table~\ref{tab:f2-completion} shows that the spherical neutral-atom reference
reduces the direct error from $-52.4$ to $0.865$ mHa after transfer to the
molecule.  Exchange remains unchanged, and its $15.87$ mHa error dominates
the uncompleted RHF result.  An exchange-only correction sets
$\Delta E_K=0$ but leaves the $0.865$ mHa direct error and does not reproduce
the complete reference Fock vectors.  The full $X_{\rm HF}$ correction
instead matches all nine occupied Fock vectors and the finite-reference RHF
energy to numerical precision.

Orbital reoptimization provides the more important robustness check.  The
unscreened, core-screened, and neutral-screened energies fall by 10.02, 4.68,
and 0.91 mHa, respectively.  All three calculations remain in the same
restricted orbital basin, but screening progressively reduces the drift from
the reference.  With $X_{\rm HF}$ included, restricted-HF calculations
starting from both the reference orbitals and several perturbed starting
points return to the reference solution.  This provides a practical test of
local stability within restricted HF; the known instability toward UHF
remains.

\subsection{Heavy-atom test:
\texorpdfstring{Cr$_2$}{Cr2}}

Cr$_2$ provides a difficult molecular test of Hartree screening and
finite-reference matching.  At $R=3.175$ bohr, we construct standard $q=5$
and $q=7$ PQS bases around the same broken-symmetry Cartesian cc-pV5Z UHF
reference, with 24 occupied orbitals of each spin and
$\langle S^2\rangle=4.866$.  Its 448 raw Cartesian AOs reduce to a retained
rank of 436.  The
$q=5$ working basis contains 1795 PQS functions and 138 residual cc-pV5Z
functions with $l\leq2$, giving $N=1933$; the corresponding $q=7$ dimensions
are 6915, 138, and 7053.  Thus neither working basis contains the complete
436-dimensional Gaussian space.  We first project the reference determinant
into each working basis.  PySCF then evaluates $hQ$, $JQ$, $KQ$, and $FQ$ for
this captured determinant; their projections into the same working basis
define the component comparisons and $X_{\rm HF}$.  For each basis we compare
the unscreened Hamiltonian with versions screened by two fitted spherical
atomic Hartree fields.  The core reference contains
the 18 electrons of Cr$^{6+}$.  The neutral reference adds the six-electron
$3d^5 4s^1$ density, with the five $d$ orientations averaged to give a
spherical atomic charge.  Each field is placed on the two atoms without
molecular refitting.  For each Hartree treatment we construct its own
spin-resolved $X_{\rm HF}$.  The difference
between the gausslet and PySCF physical one-particle energies at the captured
determinant,
$\Delta E_h=E_h^{\rm model}[\Phi_0]-E_h^{\rm ref}[\Phi_0]$, is common to all
three Hartree treatments: 6.59 mHa at $q=5$ and 0.501 mHa at $q=7$;
adding it to $\Delta E_J+\Delta E_K$ gives the fixed UHF errors in
Table~\ref{tab:cr2}.

\begin{table*}[t]
\caption{\label{tab:cr2}Core and neutral-atom screening for Cr$_2$ at
$R=3.175$ bohr.  Errors are relative to the captured cc-pV5Z UHF determinant
and are in mHa except where noted.  Arrows give the fixed-determinant value
followed by the outcome of HF optimization.  The entries marked with a $\ast$ denote large negative energy errors where the accurate transferred state was unstable and relaxation transitioned to a false state; energy errors in these cases are approximate.
$n_G$ is the residual-Gaussian occupation after relaxation with
$X_{\rm HF}$.}
\begin{ruledtabular}
\begin{tabular}{rrlrrrrr}
$q$ & $N$ & Hartree treatment & $\Delta E_J$
& $\Delta E_K$
& $\Delta E_{\rm HF}$
& $\Delta E_{\rm HF}^{+X_{\rm HF}}$
& $n_G^{+X_{\rm HF}}$\\
\hline
5 & 1933 & unscreened       & $-2058$ & $162.4$  & $-1.889~{\rm Ha}\rightarrow-23.44^*~{\rm Ha}$ & $0\rightarrow -20.67^*~{\rm Ha}$ & ---\\
5 & 1933 & Cr$^{6+}$ core  & $-54.6$ & $162.4$  & $114.4\rightarrow95.2$  & $0\rightarrow-0.163$ & 0.0773\\
5 & 1933 & neutral Cr      & $-28.6$ & $162.4$  & $140.5\rightarrow122.6$ & $0\rightarrow-0.163$ & 0.0773\\
7 & 7053 & unscreened       & $-42.0$ & $5.56$ & $-35.93\rightarrow-582.7^*$ & $0\rightarrow -482.0^*$ & ---\\
7 & 7053 & Cr$^{6+}$ core  & $-5.15$ & $5.56$ & $0.903\rightarrow0.153$ & $0\rightarrow-0.024$ & 0.00283\\
7 & 7053 & neutral Cr      & $-0.430$ & $5.56$ & $5.63\rightarrow5.15$  & $0\rightarrow-0.024$ & 0.00283\\
\end{tabular}
\end{ruledtabular}
\end{table*}

Screening changes the direct field but leaves exchange unchanged.  Core
screening removes most of the large direct error at both orders, and the
neutral reference reduces it further.  At $q=7$, for example, the small
core-screened fixed error results from cancellation between a $-5.15$ mHa
direct error and a $5.56$ mHa exchange error.  Neutral screening reduces the
direct error to $-0.43$ mHa, but exposes the exchange error and therefore
worsens the uncompleted total.  A spherical neutral atom is thus not
uniformly better for a molecule with an open $d$ shell.

The relaxation step in Table~\ref{tab:cr2} reveals a more serious consequence
of using the unscreened diagonal interaction.  We begin with the conventional
Gaussian-basis UHF determinant represented in the gausslet working basis, and
then optimize the orbitals using the unscreened gausslet Hamiltonian.  If this
Hamiltonian were accurate near the reference state, the optimization would
make only a modest change.  Instead, it drives the orbitals into spurious
low-energy states with large occupation of the residual Gaussian functions, which have interactions represented by the less accurate matched width Gaussian approximation\cite{QiuWhite2021}, rather than IDA.
For $q=5$, the energy falls by 21.55 Ha from its value at the
starting determinant, while the residual-Gaussian occupation grows to 7.71
electrons; the resulting state is nearly closed shell and is not the intended
Cr$_2$ state.  Even $q=7$ fails: its energy falls by 0.547 Ha and the
residual-Gaussian occupation reaches 1.44 electrons.

Adding $X_{\rm HF}$ to the unscreened Hamiltonian does not cure the failure.
It reproduces the transferred energy and occupied Fock vectors, but this correction, tied closely to the occupied space, is inadequate.  For both
corrected unscreened Hamiltonians, the transferred state was unstable and optimization again reaches low energy false states.   In contrast, core and neutral screening
preserve the intended Cr$_2$ state.  At $q=5$, screening alone still permits
an 18--19 mHa energy lowering, and the optimized energy remains about 0.1 Ha
above the Gaussian reference.  The corresponding $q=7$ relaxation is less
than 1 mHa.

With screening in place, the corresponding $X_{\rm HF}$ correction is able to finish the job, making HF truly accurate and even more stable. Core only and neutral screening both work well with roughly the same accuracy. The unrelaxed state matches
the energy and occupied Fock vectors of the captured finite Gaussian-basis
reference.  After relaxation,
the core- and neutral-screened corrected states are very close in energy and
remain within 0.2 mHa of the original reference. All eight tested occupied–unoccupied perturbations at each \(q\) return to the corrected states.   

Why are the post-relaxation energies so close? The separately constructed \(X_{\rm HF}\) corrections give the two Hamiltonians the same energy and orbital gradient at the reference state. They then differ only through the unoccupied-space part of the screening correction from the small IDA error of the relatively smooth six-electron valence density. This difference changes the curvature of an already very small orbital relaxation rather than contributing directly to the reference energy. In these calculations the resulting relaxed energies differ by less than \(10^{-3}\) mHa at both orders.

The $q=5$ basis uses about one quarter as many functions as $q=7$, but its
residual-Gaussian occupation is 0.077 rather than 0.0028 electrons.  It is an
accurate, locally robust hybrid representation of the selected mean-field
state, while $q=7$ represents that state more naturally in its PQS sector.
This does not establish the $q=5$ Hamiltonian's accuracy for correlated
states.

\section{Discussion and outlook}
\label{sec:discussion}

We have introduced reference-density Hartree screening as a way to improve
the integral diagonal approximation in nested gausslet Hamiltonians.  Near a
many-electron nucleus, the large nuclear attraction and the Hartree repulsion
from the core electrons should cancel before an approximation is made.  We
accomplish this by representing the Hartree field of a reference density
accurately and applying the diagonal approximation only to density
fluctuations about that reference.  The cancellation then occurs in the
one-particle Hamiltonian, while all electrons remain active and the residual
electron--electron interaction retains its two-index form.

The calculations from He through F$_2$ using the recently developed nested PQS construction\cite{WhitePQS2026} show both the accuracy and the range
of these corrections.  He provides a simple check, although it is a special
case because its Hartree self-interaction is canceled by exchange.  Ne is
a clearer closed-shell test: neutral-atom screening nearly removes the direct
error for both shell sizes $q=5$ and $q=7$, and adding $X_{\rm HF}$ gives an accurate
mean-field result already at $q=5$.  For open-shell F, a spherical ensemble
density gives an orientation-independent atomic reference.  The same fitted
atomic field transfers without refitting to F$_2$ and removes nearly all of
its direct error.  Thus a compact one-center field can incorporate the main
local screening physics without constructing a molecular Hartree field.

The choice of reference density is nevertheless important.  An ionic core
provides a robust minimal reference, while a neutral atom also includes the
valence Hartree field when its spherical density is a good description of
the molecular environment.  The neutral references work especially well for
Ne and F$_2$.  For Cr$_2$, neutral screening improves the direct energy but
is not uniformly better than core screening once the orbitals are relaxed.
The spherical free-atom $3d^5 4s^1$ density is only an approximate
description of the molecular Hartree field, and exchange remains unchanged.
Direct and exchange errors must therefore be considered
separately rather than inferred from their sum.

Hartree screening is the most important and most physically motivated correction developed
here.  The $X_{\rm HF}$ correction has a different purpose. Given a reference conventional basis Hartree--Fock calculation,
it gives a
low-rank one-particle operator that reproduces that calculation's energy and
occupied Fock vectors.  In this way the remaining accuracy of a high-quality
Gaussian-basis mean-field calculation can be transferred to the gausslet
Hamiltonian after the direct screening problem has been removed. Its limitations are that  it is state-specific, and it leaves the
orthogonal--orthogonal Fock block of the underlying model unchanged and need
not repair an instability in that space.

Cr$_2$ is the most demanding practical test.  Starting from the accurate
Gaussian-basis HF state, relaxation with neither type of correction to the hybrid gausslet
Hamiltonian moves to spurious
states dominated by the residual Gaussians, lowering the energy by 21.55 Ha
for $q=5$ and 0.547 Ha for $q=7$.  Applying $X_{\rm HF}$ \emph{without screening} also matches the selected energy and
occupied Fock vectors, but relaxation again moves  to false states at least 20.67 and 0.482 Ha
below the reference.  Both core and neutral screening prevent this failure
and preserve the intended state under relaxation in both bases. After screening and
$X_{\rm HF}$, the
1933-function $q=5$ basis reproduces the selected mean-field state to sub-mHa
accuracy and is locally stable.  This basis is only 4.4 times the rank-436
cc-pV5Z reference space, while retaining a two-index rather than a four-index
interaction.  The $q=7$ basis has much smaller residual-Gaussian occupation, and makes a natural next higher accuracy basis.

It remains to be established how accurate these relatively small diagonal bases are for the treatment of correlation,
after their
mean-field errors have been removed.  Neither screening nor $X_{\rm HF}$
supplies the spatial resolution needed for the electron--electron cusp, and
$X_{\rm HF}$ is constructed only from its action on the reference
determinant.  Nevertheless, compact gausslet bases with two-index
interactions, such as $q=5$, should be valuable starting points for correlated
calculations.  Calculations at larger $q$ can then test convergence.  More
broadly, the present results show that the dominant
nuclear--Hartree cancellation can be built accurately into a gausslet
Hamiltonian without sacrificing the two-index interaction that makes
gausslets attractive for large correlated calculations.

\acknowledgments
I thank Sandeep Sharma for valuable discussions.  This work was supported by the U.S. National Science Foundation under Grant
DMR-2412638.

\section*{AUTHOR DECLARATIONS}

\subsection*{Conflict of Interest}
The author has no conflicts to disclose.

\subsection*{Author Contributions}
\textit{Steven R. White}: Conceptualization; Methodology; Software;
Validation; Formal analysis; Investigation; Data curation; Visualization;
Writing---original draft; Writing---review and editing; Funding acquisition.

\section*{DATA AVAILABILITY}
The numerical data that support the findings of this study are available from
the corresponding author upon reasonable request.  The basis-construction and
screened-Hartree software is available in the open-source
\texttt{GaussletBases.jl} repository at
\url{https://github.com/srwhite59/GaussletBases.jl}.

\bibliographystyle{aipnum4-2}
\bibliography{ref}

\end{document}